\documentclass[twocolappendix,twocolumn]{aastex63}

\definecolor{dark-red}{rgb}{0.4,0.15,0.15}
\definecolor{dark-blue}{rgb}{0.15,0.15,0.6}
\definecolor{medium-blue}{rgb}{0,0,0.5}
\hypersetup{
    colorlinks, linkcolor={dark-red},
    citecolor={dark-blue}, urlcolor={medium-blue}
}

\shorttitle{SNIascore}
\shortauthors{Fremling et al.}
 \watermark{draft}
\graphicspath{{./}{figures/}}

\newcommand{\SNIascore}{\texttt{SNIascore}}
\newcommand{\SNIascoreerr}{\texttt{$\sigma_\mathrm{SNIascore}$}}
\newcommand{\zerr}{\texttt{$\sigma_\mathrm{z}$}}

\newcommand{\totalspectra}{5550 }
\newcommand{\totaltransients}{3463 }
\newcommand{\validationspectra}{1016 }
\newcommand{\validationtransients}{648 }
\newcommand{\testingspectra}{1011 }
\newcommand{\testingtransients}{632 }

\newcommand{\totalIaspectra}{1526 }
\newcommand{\totalIa}{1090 }
\newcommand{\totalNotIaspectra}{1997 }
\newcommand{\totalNotIa}{1121 }

\newcommand{\zpredictionspectra}{891 }
\newcommand{\zpredictiontransients}{630 }

\newcommand{\augmentspectra}{12810 }

\begin{document}

\title{SNIascore: Deep Learning Classification of Low-Resolution Supernova Spectra}

\correspondingauthor{C.~Fremling}
\email{fremling@caltech.edu}

\author[0000-0002-4223-103X]{Christoffer~Fremling}
\affiliation{Division of Physics, Mathematics, and Astronomy, California Institute of Technology, Pasadena, CA 91125, USA}

\author[0000-0002-9364-5419]{Xander~J.~Hall}
\affiliation{Division of Physics, Mathematics, and Astronomy, California Institute of Technology, Pasadena, CA 91125, USA}

\author[0000-0002-8262-2924]{Michael W. Coughlin}
\affiliation{School of Physics and Astronomy, University of Minnesota, Minneapolis, Minnesota 55455, USA}

\author{Aishwarya S. Dahiwale}
\affiliation{Division of Physics, Mathematics, and Astronomy, California Institute of Technology, Pasadena, CA 91125, USA}

\author[0000-0001-5060-8733]{Dmitry~A.~Duev}
\affiliation{Division of Physics, Mathematics, and Astronomy, California Institute of Technology, Pasadena, CA 91125, USA}

\author{Matthew J. Graham}
\affiliation{Division of Physics, Mathematics, and Astronomy, California Institute of Technology, Pasadena, CA 91125, USA} 

\author[0000-0002-5619-4938]{Mansi M. Kasliwal}
\affiliation{Division of Physics, Mathematics, and Astronomy, California Institute of Technology, Pasadena, CA 91125, USA}

\author[0000-0002-7252-3877]{Erik~C.~Kool}
\affiliation{The Oskar Klein Centre, Department of Astronomy, Stockholm University, AlbaNova, SE-10691 Stockholm, Sweden} 

\author[0000-0003-2242-0244]{Ashish~A.~Mahabal}
\affiliation{Division of Physics, Mathematics, and Astronomy, California Institute of Technology, Pasadena, CA 91125, USA}
\affiliation{Center for Data Driven Discovery, California Institute of Technology, Pasadena, CA 91125, USA}

\author[0000-0001-9515-478X]{Adam~A.~Miller}
\affiliation{Center for Interdisciplinary Exploration and Research in Astrophysics and Department of Physics and Astronomy, Northwestern University, 1800 Sherman Ave, Evanston, IL 60201, USA}
\affiliation{The Adler Planetarium, Chicago, IL 60605, USA}

\author[0000-0002-0466-1119]{James D. Neill}
\affiliation{Division of Physics, Mathematics, and Astronomy, California Institute of Technology, Pasadena, CA 91125, USA}

\author[0000-0001-8472-1996]{Daniel~A.~Perley}
\affiliation{Astrophysics Research Institute, Liverpool John Moores University, Liverpool Science Park, 146 Brownlow Hill, Liverpool L35RF, UK}

\author[0000-0002-8121-2560]{Mickael Rigault}
\affiliation{Univ Lyon, Univ Claude Bernard Lyon 1, CNRS, IP2I Lyon / IN2P3, IMR 5822, F-69622, Villeurbanne, France}

\author[0000-0002-6099-7565]{Philippe Rosnet}
\affiliation{Universit\'e Clermont Auvergne, CNRS/IN2P3, Laboratoire de Physique de Clermont, F-63000 Clermont-Ferrand, France}

\author[0000-0001-7648-4142]{Ben Rusholme}
\affiliation{IPAC, California Institute of Technology, 1200 E. California Blvd, Pasadena, CA 91125, USA}

\author[0000-0003-4531-1745]{Yashvi~Sharma}
\affiliation{Division of Physics, Mathematics, and Astronomy, California Institute of Technology, Pasadena, CA 91125, USA}

\author[0000-0002-1486-3582]{Kyung Min Shin}
\affiliation{Division of Physics, Mathematics, and Astronomy, California Institute of Technology, Pasadena, CA 91125, USA}

\author[0000-0003-4401-0430]{David L. Shupe}
\affiliation{IPAC, California Institute of Technology, 1200 E. California Blvd, Pasadena, CA 91125, USA}

\author[0000-0003-1546-6615]{Jesper~Sollerman}
\affiliation{The Oskar Klein Centre, Department of Astronomy, Stockholm University, AlbaNova, SE-10691 Stockholm, Sweden}

\author[0000-0001-8018-5348]{Richard S. Walters}
\affiliation{Division of Physics, Mathematics, and Astronomy, California Institute of Technology, Pasadena, CA 91125, USA} 

\author[0000-0001-5390-8563]{S. R. Kulkarni}
\affiliation{Division of Physics, Mathematics, and Astronomy, California Institute of Technology, Pasadena, CA 91125, USA}

\begin{abstract}
We present \SNIascore, a deep-learning based method for spectroscopic classification of thermonuclear supernovae (SNe Ia) based on very low-resolution (R$\sim100$) data. The goal of \SNIascore\, is fully automated classification of SNe Ia with a very low false-positive rate (FPR) so that human intervention can be greatly reduced in large-scale SN classification efforts, such as that undertaken by the public Zwicky Transient Facility (ZTF) Bright Transient Survey (BTS). We utilize a recurrent neural network (RNN) architecture with a combination of bidirectional long short-term memory and gated recurrent unit layers. \SNIascore\, achieves a $<0.6\%$ FPR while classifying up to $90\%$ of the low-resolution SN Ia spectra obtained by the BTS. \SNIascore\, simultaneously performs binary classification and predicts the redshifts of secure SNe Ia via regression (with a typical uncertainty of $<0.005$ in the range from $z=0.01$ to $z=0.12$). For the magnitude-limited ZTF BTS survey ($\approx70\%$ SNe Ia), deploying \SNIascore\, reduces the amount of spectra in need of human classification or confirmation by $\approx60\%$. Furthermore, \SNIascore\, allows SN Ia classifications to be automatically announced in real-time to the public immediately following a finished observation during the night.

\end{abstract}



\section{Introduction} \label{sec:intro}

Modern time-domain surveys, such as the Zwicky Transient Facility \citep[ZTF;][]{Bellm2019,Bellm2019b,Graham2019,Masci2019,Dekany20}, the All-Sky Automated Survey for Supernovae \citep[ASAS-SN;][]{Shappee2014} and the Asteroid Terrestrial Last-Alert System \citep[ATLAS;][]{Tonry2018}, are now finding tens of thousands of transients every year.

However, without spectroscopic classifications these discoveries are of limited value \citep{Kulkarni2020}. The ZTF Bright Transient Survey (BTS; \citealp{Fremling2020,Perley2020rcf}) is addressing this through the deployment of a fully automated very-low-resolution spectrograph, the Spectral Energy Distribution Machine (SEDM; \citealt{Blagorodnova2018,Rigault2019}) mounted on the Palomar 60-inch telescope. SEDM is capable of obtaining spectra of several thousands of transients per year in the magnitude range between 18 and 19 mag. Currently, the goal of the BTS is to maintain spectroscopic classification completeness for all extragalactic transients detected by the public ZTF survey that become brighter than 18.5 mag ($\sim 1000$ SNe per year; \citealp{Perley2020rcf}).

The classifications from the BTS are made public on a daily basis via the Transient Name Server (TNS\footnote{\url{https://wis-tns.org}}). These classifications have up until now been based on manual matching of observed spectra to spectral templates using mainly the \texttt{SuperNova} \texttt{IDentification} (\texttt{SNID}; \citealt{Blondin2007}) code, along with careful inspection of each obtained spectrum. This makes classification of thousands of SNe a very time-consuming endeavor.

Due to their inherent brightness, the majority of the extragalactic transients discovered by a magnitude limited survey will be thermonuclear supernovae\footnote{In the ZTF BTS $\approx72.5\%$ of the extragalactic transients detected are SNe Ia \citep{Perley2020rcf}.} (SNe Ia). Here we present \SNIascore, a deep-learning based method optimized to identify SNe Ia using SEDM spectra and determine their redshifts without any human interaction. The intended use case for \SNIascore\ is to provide live spectroscopic classification of SNe Ia during the night when SEDM is observing for the BTS.

\begin{deluxetable}{lrrrrr}
\tabletypesize{\footnotesize}
\tablecolumns{5}
\tablewidth{0pt}
\tablecaption{Data summary \label{tab:data}}
\tablehead{
\colhead{Class} & \colhead{} & \colhead{All} & \colhead{Training} & \colhead{Validation} & \colhead{Testing} 
}
\startdata
\textbf{SNIa}           & & 2619 & 1526 &  607 &  486 \\
\textbf{NotSNIa}	    & & 2931 & 1997 &  409 &  525 \\
\phm{sss}H-rich CC SN   & & 1285 &  751 &  312 &  222 \\
\phm{sss}H-poor CC SN   & & 585  &  393 &   94 &  98 \\ 
\phm{sss}TDE	        & & 37   &   35 &    0 &  2 \\
\phm{sss}CV	            & & 325  &  284 &    0 &  41 \\
\phm{sss}Other          & & 699  &  534 &    3 &  162 \\
\enddata
\tablecomments{All numbers refer to the number of unique spectra in each class. The ``Other'' class includes any spectra that does not fit the other classes, including galaxy spectra, active galactic nucleus (AGN) spectra, and spectra that BTS has been unable to classify as belonging to any known transient class.} 
\end{deluxetable}

\SNIascore\, is based on a recurrent neural network (RNN) architecture (see \citealp{RNNreview} for a review) with a combination of bidirectional long short-term memory (BiLSTM) and gated recurrent unit (GRU) layers. \SNIascore\, is able to classify $>80\%$ of the SN Ia spectra that are observed by SEDM for the ZTF BTS with a false-positive rate (FPR) of $<1\%$. 

\SNIascore\, was trained on SEDM data obtained by the ZTF BTS between 2018 March and 2020 March, and validated and optimized with the BTS dataset published by \cite{Fremling2020}. A final test run is performed using data obtained between 2020 April and 2020 August which were neither part of the training nor validation datasets. Our datasets are described in detail in Section~\ref{sec:data}. The network architecture is described in Section~\ref{sec:arch}, and the training and optimization procedure used for \SNIascore\, is described in Section~\ref{sec:training}. The performance is evaluated and compared to SNID and the previously published deep-learning method DASH \citep{MuthukrishnaDASH} in Section~\ref{sec:testing}. A discussion on the most likely false positives can be found in Section~\ref{sec:discussion}. The implementation of \SNIascore\, as part of the ZTF BTS is described in Section~\ref{sec:implementation}. Future development possibilities are also discussed in this section.

\begin{figure}[t!]
\hspace{-0.35cm}\includegraphics[width=0.51\textwidth]{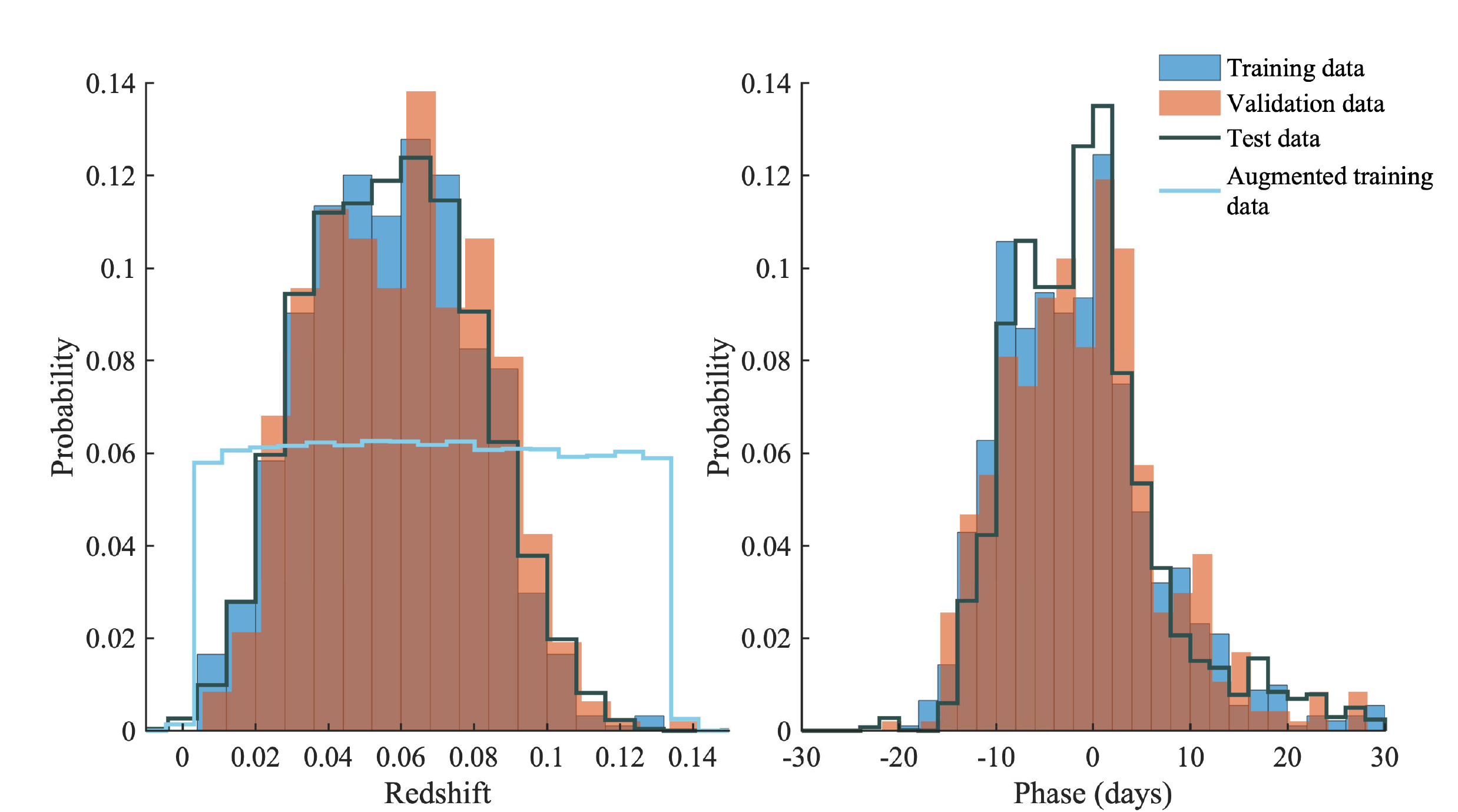}
\caption{Redshift (left) and spectral phase (right) distributions for SNe Ia in the unaugmented \SNIascore\, training (blue) and validation (red) datasets used to optimize \SNIascore\, for classification. The SN Ia distributions for the testing dataset are shown in black. The blue line in the left panel shows the redshift distribution of the training set after augmentation (Section~\ref{sec:augment}), which is used to train \SNIascore\, for redshift regression. The phase is relative to the time of maximum light in the $g$ or $r$ band, depending on which band is brighter.} 
\label{fig:sample}
\end{figure}

\section{Datasets} \label{sec:data}
Our data consist of SEDM spectra of transients detected by ZTF, which were followed up and classified by the BTS \citep{Fremling2020,Perley2020rcf} using the \texttt{GROWTH Marshal} \citep{Kasliwal2019GM}. These spectra are of low resolution (R$\sim100$), and cover a typical wavelength range of 3800~\AA\, to 9150~\AA\, within 209 wavelength bins (see Section~\ref{sec:preprocessing}). Although the BTS focuses on extragalactic transients, some spectra eventually turn out to be of Galactic sources (e.g., cataclysmic variables; CVs). We include both spectra of Galactic and extragalactic transients in our datasets. The dataset we use here contains \totalspectra SEDM spectra of \totaltransients individual transients obtained between 2018 March and 2020 August. A breakdown of the various classes of transients included in our dataset can be found in Table~\ref{tab:data}. We split our full dataset in three parts: training data, validation and optimization data, and final performance testing data. 

For the purpose of training \SNIascore\, (Section~\ref{sec:training}) we group the data into two classes: real SNe Ia (\textbf{SNIa}; \totalIaspectra spectra of \totalIa SNe Ia), and everything else (\textbf{NotSNIa}; \totalNotIaspectra spectra of \totalNotIa transients). The training data were collected between 2018 March 7 and 2020 March 1, but we exclude any data that are part of our validation set which was also collected during 2018.

To validate and optimize \SNIascore\ (Section~\ref{sec:training}) we use \validationspectra spectra of \validationtransients SNe which are part of the BTS sample from 2018 published in \cite{Fremling2020} (hereafter the BTS18 sample). The BTS18 sample was chosen for validation since the classifications and redshifts of the SNe in this sample have been carefully vetted by humans.

For final performance testing we use all BTS spectra collected between 2020 Mar 2 and 2020 Aug 5 (\testingspectra spectra of \testingtransients transients). The classifications in this testing sample are sufficiently accurate to evaluate the performance of \SNIascore, but they are subject to minor future changes as work on the BTS proceeds.\footnote{The live BTS sample, which is updated daily, can be accessed through the BTS sample explorer at \url{https://sites.astro.caltech.edu/ztf/bts/explorer.php}.}

\begin{figure}[t!]
\vspace{0.15cm}\includegraphics[width=0.46\textwidth]{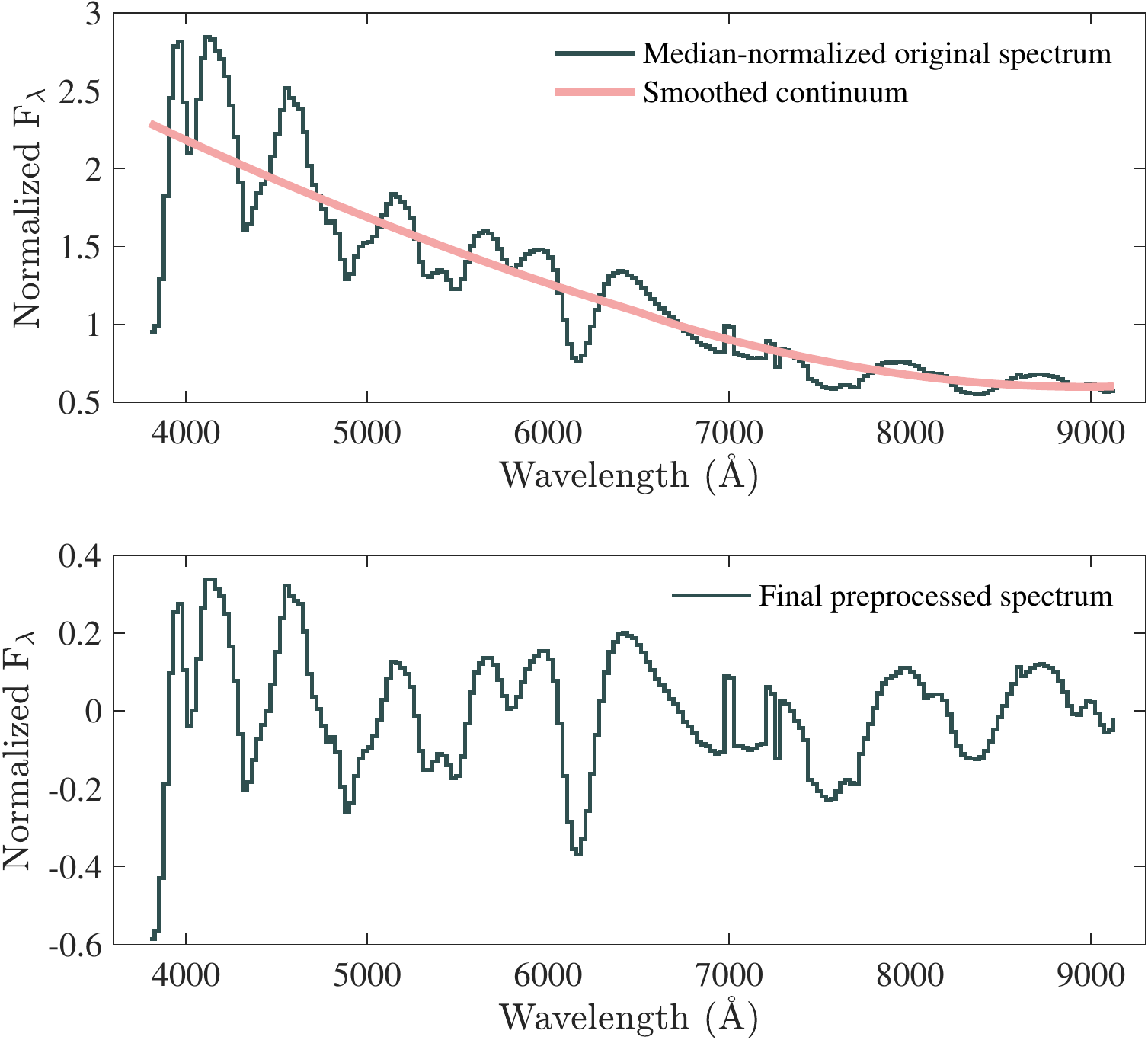}
\caption{Preprocessing procedure used in \SNIascore. The top panel shows an example SN Ia spectrum, normalized by the median value (black line). This normalized spectrum is divided by a smoothed continuum (computed with robust local polynomial regression; red line) to create the final preprocessed spectrum (black line, bottom panel), which is centered around zero by subtracting a constant value of one.}
\label{fig:preprocessing}
\end{figure}

\begin{figure*}[t!]
\includegraphics[width=1\textwidth]{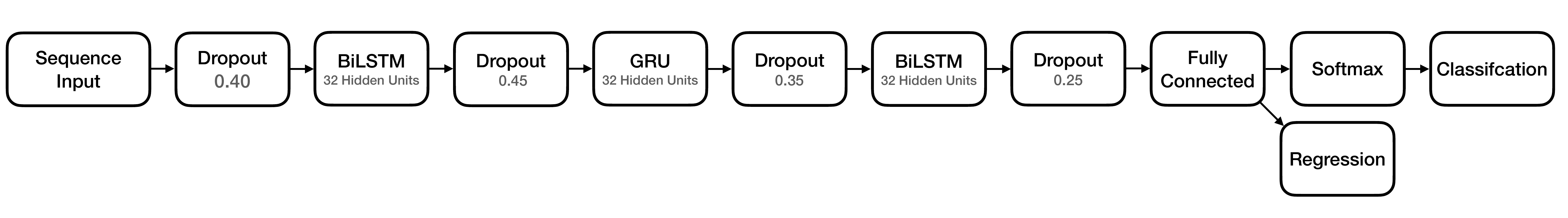}
\caption{Network architecture of \SNIascore. We utilize heavy dropout throughout the network and a combination of two BiLSTM layers surrounding one GRU layer. For regression the final softmax and classification layers are replaced by a regression layer.}
\label{fig:arch}
\end{figure*}

For training \SNIascore\, for redshift prediction we only use spectra of SNe Ia, and also impose a cut on the quality of the known redshifts. We only use SNe Ia where the redshift is known to three decimal places or more\footnote{This typically means that a spectrum of the host galaxy exists in the literature (e.g., on NED), the SEDM spectrum was of very high SNR with very good SNID template matches or clear host galaxy lines, or that we have obtained a spectrum of the SN from a telescope that offers a higher resolution compared to SEDM.}. Both redshifts derived from broad SN features and host galaxy emission lines are used. The training set for redshift prediction consists of \zpredictionspectra spectra of \zpredictiontransients SNe Ia, which is increased to \augmentspectra spectra through data augmentation (Section~\ref{sec:augment}).

The properties of the training, validation and testing datasets are illustrated in Figure~\ref{fig:sample}. The redshift range covered by our full dataset is $z=0.008$ to $z=0.126$. However, $98\%$ of our data fall within the range $z=0.01$ to $z=0.11$. The phases of our spectra with respect to maximum light determined from the lightcurves of our transients fall within $-20$~days to $+30$~days, with $97\%$ of the data within $-20$~days to $+20$~days. Thus, the range we can expect \SNIascore\, to perform reliably within is $z=0.01$ to $z=0.11$ for spectra obtained within $\pm20$ days of maximum brightness. We use the time of maximum light from the BTS Sample Explorer \citep{Perley2020rcf}, which records the brightest actual measurement in the lightcurve of each transient. This measurement can be in the $g$ or $r$ band, depending on which band is brighter.

\subsection{Training set augmentation}\label{sec:augment}
For the classification component of \SNIascore\, we can achieve excellent performance without any need for data augmentation (Section~\ref{sec:testing}). We also do not need to de-redshift our spectra as part of the pre-processing procedure that is performed before training or when classifying new data (Section~\ref{sec:preprocessing}). However, in order to use regression to predict the redshifts of SNe Ia identified by \SNIascore\, without introducing systematic bias, a weighting scheme, or data augmentation is needed. This is due to the shape of the redshift distribution of our dataset (Fig.~\ref{fig:sample}); there are very few spectra for SNe Ia at both $z<0.03$ and $z>0.09$.

We have found that a simple augmentation procedure, which equalizes the redshift distribution of the training sample, can offer excellent redshift prediction performance with negligible bias. We perform the following s.pdf: for each redshift bin of size 0.001 from zero to 0.13 we search for spectra in our unaugmented dataset with similar redshifts (within $z-0.005$ to $z+0.01$). We then randomly pick spectra from the matches, apply new redshifts by shifting the wavelengths of the original spectra to randomly selected redshifts within $\pm0.001$ of the redshift bin center (the spectra are not de-redshifted to the rest frame as part of this process to reduce loss off information at the edges of the spectra). We repeat this process until we have roughly 100 spectra in each redshift bin, for a total of \augmentspectra spectra. To reduce overfitting, and lessen the impact of repeating a small number of spectra at the edges of the redshift distribution, we introduce noise from a normal distribution to each added spectrum.\footnote{We add between zero and one standard deviation of noise. The standard deviation is estimated as the standard deviation of each spectrum after subtraction of the broad features using a heavily smoothed spectrum.} The final redshift distribution for our training data used for redshift regression is shown in Figure~\ref{fig:sample}.

\subsection{Preprocessing of spectra}\label{sec:preprocessing}
Due to small variations in the final wavelength coverage of the SEDM spectra included in our dataset we interpolate all spectra to a uniform wavelength grid, which spans the largest possible wavelength range without a need for extrapolation for any individual spectrum. This range is 3800~\AA\, to 9150~\AA\, with 209 wavelength bins of size 25.6~\AA. 

After interpolation we normalize each spectrum by division with its median value. The normalized spectrum is then divided by its continuum (a heavily smoothed version of the normalized spectrum computed with robust\footnote{Lower weights are assigned to outliers in the regression; outliers at $>6\sigma$ are assigned zero weight.} local polynomial regression using weighted linear least squares and a 2nd degree polynomial model). Finally, to center the preprocessed spectra around zero we subtract a constant value of one from each normalized and continuum divided spectrum. This process is illustrated in Figure~\ref{fig:preprocessing}.

We have found that this pre-processing procedure significantly outperforms a simple division by the median, directly dividing by a smoothed continuum, or dividing by the median and then subtracting the smoothed continuum. We have also investigated the effect of suppressing the bluest and reddest parts of the spectra (which can be very noisy in data from many instruments), as in \cite{MuthukrishnaDASH}. This decreases the overall performance and is not needed for SEDM data. 

Effectively our pre-processing procedure flattens the spectra (removes any temperature gradient or host galaxy continuum emission contribution), makes emission lines positive and absorption lines negative, and normalizes the strengths of the features. 

\section{Neural network architecture}\label{sec:arch}
For our final \SNIascore\ network that we have arrived at after following the optimization procedure described in Section~\ref{sec:training}, we have used an RNN architecture (see e.g., \citealp{RNNreview}) consisting of a combination of BiLSTM and GRU layers, with 32 hidden units in each layer (Fig.~\ref{fig:arch}). We employ significant dropout both during training and prediction. We use a dropout of $40\%$ immediately following the input layer. This is followed by a BiLSTM layer and $45\%$ dropout. After this we use a GRU layer and $35\%$ dropout and then another BiLSTM layer and $25\%$ dropout. This last BiLSTM layer is followed by a fully connected layer with two outputs, a softmax layer and a classification layer which computes the cross-entropy loss for the \textbf{SNIa} and \textbf{NotSNIa} classes.

For redshift regression the softmax and classification layers are replaced by a custom regression layer, which uses the root of the mean bias error (MBE; Eq.~\ref{eq:MBE}) squared as the forward loss function and the derivative of the mean absolute error (MAE; Eq.~\ref{eq:MAE}) as the backward loss function. We have found that this custom regression layer gives the least systematic bias and best overall performance among the loss functions we have tested (Section~\ref{sec:training}).

To estimate uncertainties when predicting both the \SNIascore\, (\SNIascoreerr) and redshift (\zerr), we use a simple Monte-Carlo (MC) method. We re-run each prediction 100 times and adopt the standard deviation of the results as the uncertainty of the prediction.  

\section{Training and optimization}\label{sec:training}
\subsection{Classification Optimization}
We have implemented the architecture for \SNIascore\, described in Section~\ref{sec:arch} using the \texttt{MATLAB} Deep Learning Toolbox$^{\mathrm{TM}}$. For training we use the adaptive learning rate optimization algorithm \texttt{Adam} \citep{ADAM} with the following settings:
\begin{itemize}
\setlength{\itemsep}{-2pt}
\item{MiniBatchSize: 256} 
\item{InitialLearnRate: 0.005} 
\item{LearnRateSchedule: none} 
\item{L2Regularization: $10^{-4}$} 
\item{GradientThreshold: 2} 
\item{Shuffle: every-epoch} 
\item{SequenceLength: longest} 
\end{itemize}
For validation and optimization we use the BTS18 dataset. To arrive at the architecture described in Section~\ref{sec:arch} and the optimal hyperparameters listed above we have followed this procedure:
\begin{itemize}
\setlength{\itemsep}{-2pt}
\item{We construct a grid of networks with an initial dropout layer following the input layer and then one to five BiLSTM layers with dropout layers between each layer.} 
\item{For each network we create a grid of subnetworks with a range of hyperparameters; dropout from 0.15 to 0.45 for each dropout layer and hidden units from 8 to 128 for each BiLSTM layer.} 
\item{We explore learning rates in the range 0.001 to 0.01 and mini-batch sizes of $2^n$ in the range 16 to 512 for each network.}
\end{itemize}
This optimization procedure showed that a constant learning rate of 0.005 and a mini-batch size of 256 seems to perform universally well on our data, regardless of the other hyperparameters. Changes to the learning rate (including introducing adaptive learning rate schemes), gradient threshold, and L2 regularization listed above have negligible impact on the final performance. As long as the network retains the ability to converge during training, it remains possible to pick an optimal epoch from the training sequence where the performance is very similar.

\begin{figure*}[ht!]
\begin{center}
\includegraphics[width=0.95\textwidth]{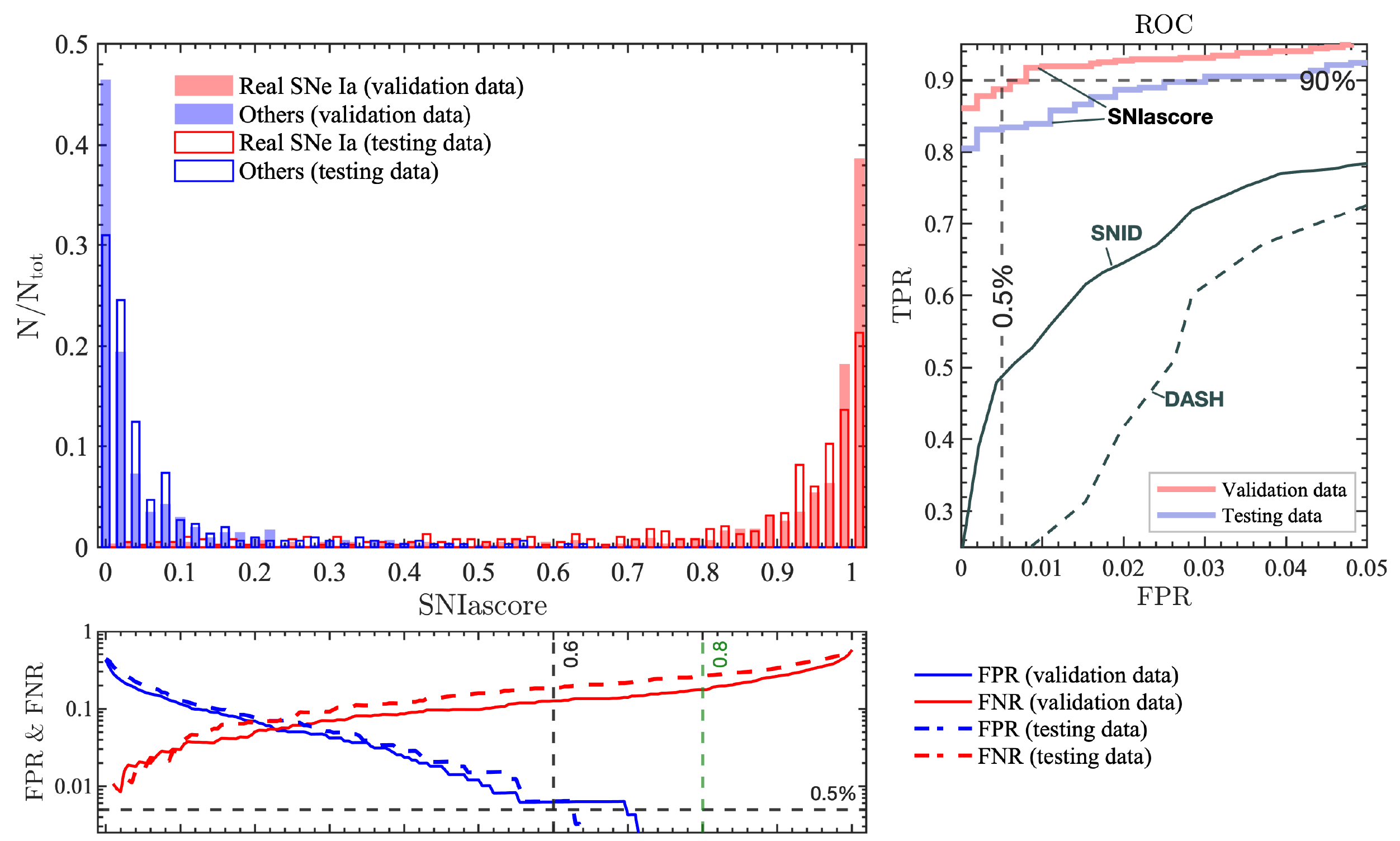}
\end{center}
\caption{Classification performance of \SNIascore\, on the BTS18 validation sample and on the testing dataset. \textit{Top left panel:} The \SNIascore\ distributions for both ``Real SNe Ia" spectra (red) and spectra of ``Other" transients (blue). The validation set is shown as solid bars and the testing dataset is shown as open bars. $N_\mathrm{tot}$ is the total for each class separately (i.e. the sum of the ``Real SNe Ia" bins add up to unity). \textit{Bottom panel:} FPR and FNR as a function of \SNIascore\ for the validation (solid red and blue lines) and for the testing (dashed red and blue lines) datasets. A cut on \SNIascore\ at 0.6 results in an FPR of roughly $0.5\%$ and TPR of $90\%$ on the validation dataset and $83\%$ on the testing dataset. A cut at a \SNIascore\ of 0.8 results in a FPR of zero on both the validation and testing datasets, with a TPR of $86\%$ and $80\%$, respectively. \textit{Right panel:} ROC curve for \SNIascore\ on the validation dataset (red line) and the testing dataset (blue line). For comparison we also show ROC curves for SNID (solid black line) and DASH (dashed black line).}
\label{fig:SNIascore}
\end{figure*}

We do not try to optimize for the required physical training time, or number of epochs required to achieve a good performance. We do not employ any stopping condition based on the loss. Instead, each network is trained well past the point of overfitting (at least 325 epochs), and we save the state of each network at every training epoch. We then choose and compare the optimal epoch (typically found between epochs 200 and 300) for every individual network based on the performance on the BTS18 validation set, according to a combination of two metrics: (i) the total number of successful classifications for a false-positive rate (FPR) of $0\%$, and (ii) the total number of successful classifications for a FPR  of $1\%$. To determine the optimal FPR and true-positive rate (TPR) of an individual network, all possible cuts on \SNIascore\ and \SNIascoreerr\ are evaluated. These metrics direct the optimization procedure towards hyperparameters and a network structure that results in a low FPR, with a high true-positive rate (TPR) being secondary. We consider FPR~$<1\%$ to be a hard requirement for autonomous spectroscopic classification. It also allows all the trained networks to be quantitatively compared based on their optimal epochs and \SNIascore\ cuts.

The architecture that resulted from this optimization procedure was three BiLSTM layers with 32 hidden units in each layer, in combination with heavy dropout (0.40, 0.45, 0.35, 0.25 for the four dropout layers from first to last). As a final step, we investigated the effect of replacing the BiLSTM layers with GRU layers in all possible combinations, and eventually arrived at our final \SNIascore\, network described in Section~\ref{sec:arch}, where the middle BiLSTM layer is replaced by a GRU layer. The effect of this is a reduced number of false positives with high \SNIascore\ values.

For our optimized classification network, Figure~\ref{fig:SNIascore} shows the \SNIascore\, distribution for known SNe Ia and other transients in the BTS18 validation sample. We find that $90\%$ of the SNe Ia in this sample are identified by \SNIascore\, with a $0.6\%$ FPR. The performance of our final \SNIascore\, network is investigated in detail and tested on independent data in Section~\ref{sec:testing}.  

\subsection{Redshift Regression Optimization}
For redshift regression we use the same optimal \SNIascore\, architecture found above with the softmax and classification layers replaced with a regression layer (Fig.~\ref{fig:arch}). However, we have found that the standard deep-learning mean square error (MSE; Eq.~\ref{eq:MSE}) or MAE loss functions result in significant redshift-dependent systematic bias, even after data augmentation to equalize the redshift distribution of the training data (Section~\ref{sec:augment}). Because of this, we perform an additional optimization step for the redshift regression network, which consists of an investigation of various forward and backward loss-functions in the final regression layer of the network.

The forward loss functions that we have investigated are the MAE, the mean percentage error (MPE; Eq.~\ref{eq:MPE}), the MSE, the MBE, and all possible combinations of these in pairs, with both the derivative of the MAE and MSE as the backwards loss. 

The MPE, MBE and the root square of these can all give good results. Differences are minor, but we have selected the root square bias error ($\sqrt{MBE^2}$) for our final network, based on optimizing (i) the percentage of SNe with redshift residuals less than 0.005 and 0.01 for the BTS18 validation sample, and (ii) how close the predictions on the full validation set adhere to a linear relationship with respect to the known redshifts ($y=ax+b$ where $a$ should be as close to 1 as possible and $b$ as close to zero as possible). Each network was trained past overfitting (at least 525 epochs) and the optimal epoch (typically found between epochs 400 and 500) was selected after the fact for comparison among the networks.

\section{Performance}\label{sec:testing}
In order to evaluate the classification performance of  \SNIascore\ we have performed comparisons to \texttt{SNID} \citep{Blondin2007} and \texttt{DASH} \citep{MuthukrishnaDASH} (Section~\ref{sec:classperf}). We also evaluate the accuracy of the associated redshift predictions for the SNe Ia that are identified by \SNIascore\, by comparison to host galaxy redshifts and redshifts determined after manual inspection with \texttt{SNID} (Section~\ref{sec:zperf}).

\begin{deluxetable}{lrrrrr}
\tabletypesize{\footnotesize}
\tablecolumns{5}
\tablewidth{0pt}
\tablecaption{Performance comparison of \SNIascore, \texttt{SNID} and \texttt{DASH} on the BTS18 validation dataset}\label{tab:cuts}
\tablehead{
\colhead{Method} & \colhead{TPR} & \colhead{FPR} & \colhead{Cuts\tablenotemark{a}}  
}
\startdata
\textbf{\SNIascore\ TPR}        &  \textbf{0.90} & 0.006 & \texttt{score}~$>0.6$, $\sigma$~$<0.3$\\
\textbf{\SNIascore\ FPR}	    &  0.86 & \textbf{0.000} & \texttt{score}~$>0.8$, $\sigma$~$<0.275$\\
\textbf{\texttt{SNID}}	    & 0.53 & 0.009 & $\texttt{rlap}>11.1$ \\
\textbf{\texttt{DASH}}	    & 0.24 & 0.009 & $\texttt{rlap}>7.1$, \texttt{score}~$>0.5$\\
\enddata
\tablenotetext{a}{\texttt{score} refers to the softmax score for both \SNIascore\ and \texttt{DASH}. For \SNIascore, $\sigma$ refers to \SNIascoreerr. }
\tablecomments{For \SNIascore\ TPR, \texttt{SNID}, and \texttt{DASH} we show the optimal TPR that can be had while maintaining FPR $<1\%$. For \SNIascore\ FPR we show the optimal TPR that can be had for FPR~$=0$.}
\end{deluxetable}

\subsection{Classifications}\label{sec:classperf}
Our goal with \SNIascore\, is low enough FPR ($<1\%$) so that the classifications do not require any human confirmation. To achieve this, using the standard value of \SNIascore~$>0.5$ to classify a spectrum as that of a SN Ia is not adequate. Instead, we evaluate all possible cuts on  \SNIascore\, and \SNIascoreerr\ to find a combination of cuts that result in the desired performance.

Receiver operating characteristic (ROC) curves for \SNIascore\ (computed by varying \SNIascore\ and \SNIascoreerr), SNID (by varying the \texttt{rlap} parameter threshold) and DASH (by varying \texttt{rlap} and the softmax score threshold) are shown in Figure~\ref{fig:SNIascore}. Based on this comparison it is clear that \SNIascore\ significantly outperforms these alternatives\footnote{Details on how \texttt{SNID} and \texttt{DASH} were used to produce the corresponding ROC curves can be found in the appendix (Section~\ref{sec:rocappendix}).}. Between $83\%$ (testing data) and $90\%$ (validation data) of the SN Ia spectra that SEDM observes for the ZTF BTS can be classified with FPR~$\approx0.5\%$. This performance is achieved for the cuts \SNIascore~$>0.6$ and \SNIascoreerr~$<0.3$.

\texttt{SNID} can achieve a very low FPR for SN Ia spectra when \texttt{rlap~$>10$}. However, only $\approx50\%$ of the spectra could potentially be automatically classified (FPR~$<1\%$) without supervision with \texttt{SNID}. The deep-learning based \texttt{DASH} only reaches a low enough FPR to enable automatic classification when applied to SEDM spectra for $24\%$ of spectra for any cuts on \texttt{rlap} and the softmax score. This is likely due to the fact that no spectra with similarly low resolution to those produced by SEDM were part of the \texttt{DASH} training data\footnote{We have investigated similar convolutional neural networks as used in \texttt{DASH} \citep{MuthukrishnaDASH}, trained on SEDM data, but we have been unable to beat the performance of our optimized RNN.}.

The cuts on \SNIascore\ and \SNIascoreerr\ mentioned above result in the highest TPR while keeping the FPR within the range we consider acceptable ($<1\%$). However, these cuts do result in some false positives, which we discuss in Section~\ref{sec:discussion}. More conservative cuts at \SNIascore~$>0.8$ and \SNIascoreerr~$<0.275$ result in zero false positives on the validation and training data. Depending on the use-case either choice of these cuts may be suitable. For live reporting of classifications to the community via TNS, the stricter cut seems appropriate (Section~\ref{sec:implementation}). We summarize our recommended \SNIascore\ cuts and the performance compared to \texttt{SNID} and \texttt{DASH} in Table~\ref{tab:cuts}. 

Finally, we note that a low \SNIascore\ does not necessarily mean that the transient being observed cannot be a SN Ia. There are multiple reasons for low scores, for example: low signal-to-noise, strong galaxy light contamination and significant contamination from cosmic rays that failed to be removed during data reduction. As such, a score that is lower than the cuts discussed above should generally be interpreted as that the spectrum cannot be used to make a confident SN Ia classification, and nothing further. Only when the \SNIascore\ is extremely low ($<0.01$), is there significant statistical power to support a conclusion that the transient associated with the respective spectrum cannot be a SN Ia. We have made no attempt at optimizing the FNR for low scores, but based on the FNR curves for the validation and testing datasets (Figure~\ref{fig:SNIascore}), \SNIascore~$<0.01$ corresponds to FNR~$\approx1\%$.

\begin{figure*}[ht!]
\begin{center}
\includegraphics[width=0.85\textwidth]{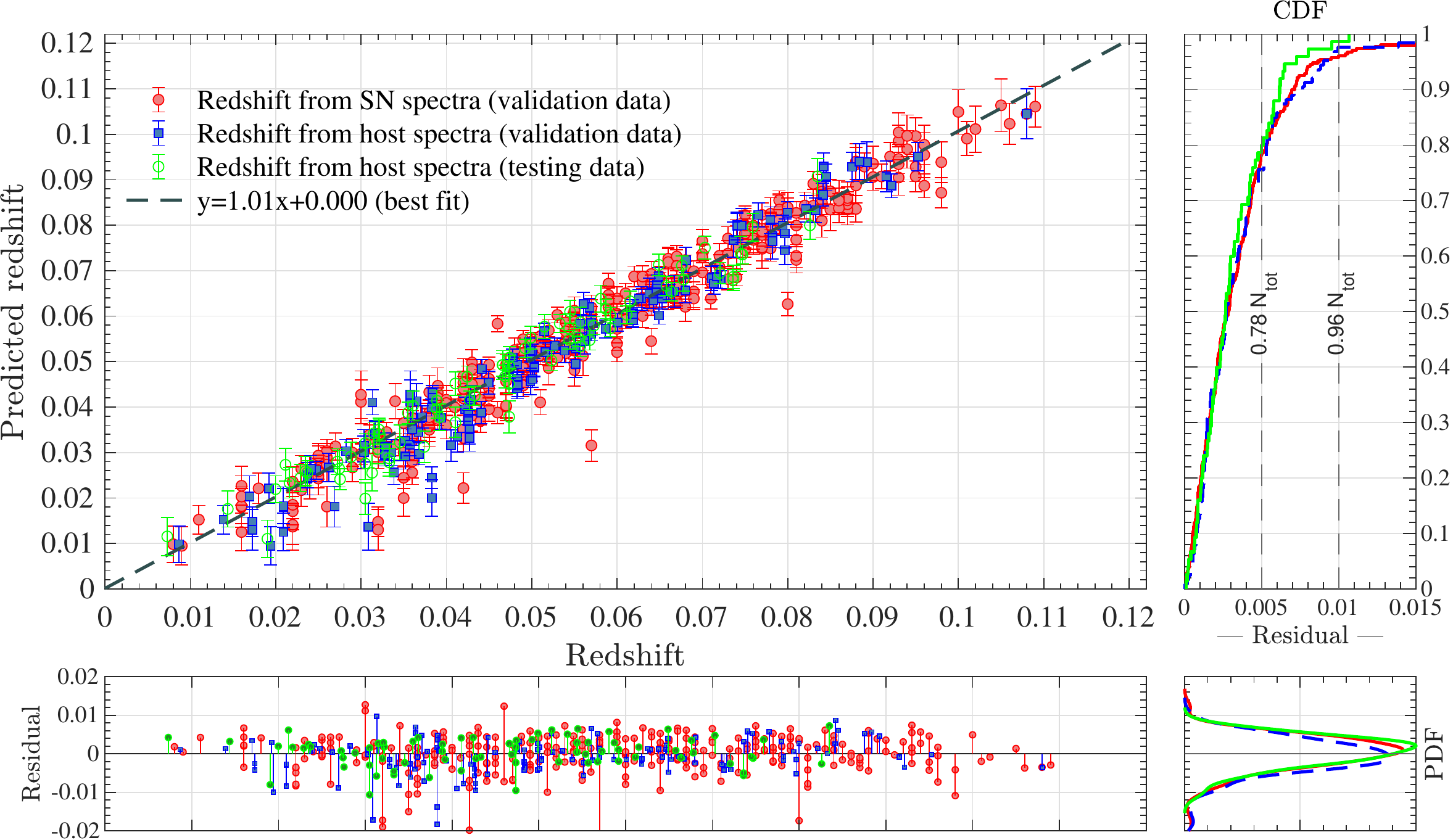}
\end{center}
\caption{Redshift regression performance for \SNIascore\ on the BTS18 validation sample (red and blue markers and lines), and on the testing dataset (green markers and lines). We find no evidence for systematic bias in the redshift predictions of \SNIascore. The best fit to the validation data is shown as a dashed black line in the main panel. The fit is consistent with $y=x$ within the uncertainties (the standard errors are $\pm0.01x$ and $\pm0.001$), and we do not find significant evidence for systematic bias in the redshift residuals (bottom left panel) of either the validation or testing datasets. Probability density functions for the residuals are shown in the bottom right panel. The typical uncertainty of a redshift estimate is $<0.005$, based on the cumulative distribution functions for the absolute value of the residuals of both the validation and testing datasets (top right panel).}
\label{fig:SNIascore_z}
\end{figure*}

\subsection{Redshifts}\label{sec:zperf}
To evaluate the accuracy of the redshifts predicted by \SNIascore\ we compare predictions for the BTS18 validation and testing datasets primarily to redshifts derived from host-galaxy spectra, and secondarily to redshifts determined from broad SN features through manual template matching using \texttt{SNID} which were performed for the BTS18 SEDM data in \cite{Fremling2020}. The performance of the \SNIascore\ redshift regression network is illustrated in Figure~\ref{fig:SNIascore_z}.

We find that the performance of our network optimized on the BTS18 validation data carries over very well to the testing dataset; we find no evidence for systematic bias in the predictions from \SNIascore\ for either dataset. The absolute value of the difference between each prediction compared to the redshift from the host-galaxy spectrum is $<0.005$ for $77\%$ of the spectra in the validation set and for $85\%$ of the spectra in the testing dataset. For $96\%$ (validation data) and $100\%$ (testing data), this difference is $<0.01$. 

We find the mean of the redshift residuals, 
$\Delta z=-0.0008$, and the standard deviation, $\sigma_z=0.0046$, for the BTS18 validation data when \SNIascore\ is compared to redshifts from host-galaxy spectra. For the testing data we find $\Delta z=0.0006$ and $\sigma_z=0.0035$, when compared to redshifts from host-galaxy spectra. When \SNIascore\ is compared to redshifts derived manually from broad SN features using \texttt{SNID} we find $\Delta z=-0.0002$ and $\sigma_z=0.0049$ (only possible for the validation data). It has been previously found that $\sigma_z=0.005$ when manual usage of \texttt{SNID} is compared to redshifts from host-galaxy spectra on the BTS18 sample \citep{Fremling2020}, which is consistent with what was found for higher resolution spectra by \cite{Blondin2007}. As such, we conclude that the performance of \SNIascore\ is consistent with what can be done with \texttt{SNID} when each result and spectrum is manually inspected. A detailed investigation of the accuracy of automatic \texttt{SNID} SN Ia redshifts from SEDM is being performed for data up until 2021 (Rigault et al., in prep.).

\section{Discussion}\label{sec:discussion}
A few spectra in both our validation and testing datasets end up as false positives with the optimal TPR cut we suggest to be used with \SNIascore\ (Table~\ref{tab:cuts}). Most of these spectra are of low-quality, or affected by strong host-contamination (and uninteresting). However, there are also a few high quality spectra. It is interesting to investigate these and the associated SNe in some detail.

In the testing set, ZTF20aatxryt (SN~2020eyj; \citealp{SN2020eyj}) stands out. Our first SEDM spectrum, taken around maximum light of the SN on 2020 April 2, is of high quality and looks consistent with a SN Ia spectrum, 
and gets an \SNIascore\ of $0.7\pm0.3$. However, it later turned out that the ejecta from this SN crashed into circumstellar material resulting in a flattening of its lightcurve and changing its spectrum into that of a SN Ibn (Kool et al., in prep.). The best matches in \texttt{SNID} are also SNe Ia with \texttt{rlap}~$>10$. However, by manually going through all the matches produced by \texttt{SNID} at \texttt{rlap}~$>5$, some SN Ic matches can also be found. Given the lightcurve evolution and the fact that SN Ia lighturve models do not match well even around maximum light before the interaction sets in, it may be more likely that this was a stripped-envelope SN. We note that the ZTF BTS officially classified this as a SN Ia, as part of our routine operations based on the SNID match to the peak spectrum \citep{ZTF20aatxryt}. Only later did we realize that the object was unusual.

In the validation set, ZTF18aaxmhvk (SN~2018cne; \citealp{SN2018cne}) stands out (\SNIascore~$=0.71\pm0.17$). Our SEDM spectrum of this source, taken around maximum light on 2018 June 14, is of high quality, and is also best matched to SNe Ia in \texttt{SNID} with very high \texttt{rlap} values. 
However, the lightcurve of ZTF18aaxmhvk is not consistent with a normal SN Ia. Based on this, and the fact that matches to SNe Ic can also be found in \texttt{SNID}, the BTS officially classified this as a SN Ic \citep{ZTF18aaxmhvk}. However, it could also have been a peculiar SN Ia. The SEDM spectrum is the only spectrum available for this source; we lack higher resolution spectra and do not have any late-time spectra, which could have provided a conclusive answer. 

In conclusion, the main contaminants that may produce high \SNIascore\ values appear to be stripped-envelope SNe, and in particular events with SN Ic-like spectra. The BTS has been classifying roughly 30 SNe Ic per year \citep{Perley2020rcf}. Thus, since there is only one SN Ic spectrum that gives a high \SNIascore\ value in each of the validation and testing datasets, it appears that only in rare cases do SNe Ic cause confusion for the BTS (which probes $z<0.1$ for core-collapse SNe). Furthermore, if the lightcurves of these SNe are taken into account it appears relatively straightforward to remove such objects from any sample of SNe Ia that is required to be devoid of both core-collapse SNe and highly peculiar SNe Ia\footnote{We note that with the low resolution of SEDM, it is not possible to spectroscopically separate between the various SN Ia subtypes, except in the most clear-cut cases (see \citealp{Fremling2020}).}. 

For future versions of \SNIascore\ we plan to add the option to include lightcurve information as input. Looking at objects that change \SNIascore\ significantly when the lightcurve is included or excluded may turn out to be a way to identify rare events that would warrant further followup from larger facilities. However, the intended use case of SNIascore is primarily to provide live classifications when SEDM is observing. As such, a potential limitation is the fact that we do not necessarily have much lightcurve information available at the time of the spectral observation by SEDM. A significant fraction of our spectra are observed before maximum light (Fig.~\ref{fig:sample}), and full lightcurves cannot be leveraged.

Some work on machine learning methods for early photometric classification has been done (e.g., RAPID; \citealp{RAPID}). However, we have not been able to reproduce the expected performance in \cite{RAPID} on live ZTF data. Work on lightcurve classifiers trained on ZTF data is in progress, which we plan to eventually combine with \SNIascore, and the redshift predictions that we already produce. Work on a deep-learning classifier capable of distinguishing between more SN subtypes while maintaining a comparably low FPR 
is also ongoing (Sharma et al., in prep.).

\vspace{1cm}

\section{Implementation}\label{sec:implementation}
Starting from 2021 April 15 we have fully deployed \SNIascore\, as a part of the automated ZTF SEDM pipeline \citep{astronote_SNIascore}. This includes automatic reporting of confident SN Ia classifications and redshift estimates to TNS. With this, we have achieved automation all the way from (spectroscopic) observation to data reduction and spectroscopic classification for the first time. 

The first fully automated SEDM SN Ia classification was sent for SN~2021ijb \citep{SN2021ijb_report}. This report was sent within roughly 10 minutes after the finished SEDM exposure. With automated ZTF candidate vetting and automatic triggering of SEDM, which is currently possible through \texttt{AMPEL} \citep{Nordin2019} and the \texttt{Fritz} marshal \citep{Skyportal2019,Duev2019rb}, a fully automated imaging (ZTF) and spectroscopic (SEDM) survey is now becoming realistic.

Initially for our TNS reports, we are using a very conservative \SNIascore\ threshold of 0.9. Based on the performance evaluation in Section~\ref{sec:classperf}, we plan to relax this to a cut on \SNIascore\ of $>0.8$ combined with a cut on \SNIascoreerr\ of $<0.275$ over the next few months. This will eliminate the need for human interaction in the classification process for the majority of BTS SNe Ia ($\approx80\%$ based on our testing data or $\approx90\%$ based on our validation data), which translates to $>50\%$ of BTS SNe.

We do not find significant variations in the FPR as a function of time, based on the data currently available to us. However, we note that the TPR (e.g., Table~\ref{tab:cuts}) and the numbers above are subject to changing observing conditions and the condition of the telescope and instrument (low SNR results in low \SNIascore). In 2020 it has been impossible to realuminize the primary mirror of the Palomar 60-inch telescope, and our testing dataset is affected by decreased overall throughput (which we currently attempt to compensate for by increased exposure times). It is possible that an increased number of SN Ia spectra can be classified when the mirror has been realuminized. Work on improvements to the SEDM data reduction pipeline is also ongoing (e.g., more effective cosmic-ray rejection and host-galaxy background subtraction).


\acknowledgments
SED Machine is based upon work supported by the National Science Foundation under Grant No. 1106171. Based on observations obtained with the Samuel Oschin Telescope 48-inch and the 60-inch Telescope at the Palomar Observatory as part of the Zwicky Transient Facility project. ZTF is supported by the National Science Foundation under Grant No. AST-1440341 and a collaboration including Caltech, IPAC, the Weizmann Institute for Science, the Oskar Klein Center at Stockholm University, the University of Maryland, the University of Washington, Deutsches Elektronen- Synchrotron and Humboldt University, Los Alamos National Laboratories, the TANGO Consortium of Taiwan, the University of Wisconsin at Milwaukee, and Lawrence Berkeley National Laboratories. Operations are conducted by COO, IPAC, and UW. This work was supported by the GROWTH project funded by the National Science Foundation under PIRE Grant No 1545949. The Oskar Klein Centre is funded by the Swedish Research Council. C.F.~gratefully acknowledges support of his research by the Heising-Simons Foundation (\#2018-0907). MR has received funding from the European Research Council (ERC) under the European Union's Horizon 2020 research and innovation programme (grant agreement n$^{\circ}$759194 - USNAC). M.~W.~C acknowledges support from the National Science Foundation with grant number PHY-2010970.


%

\facilities{P48, P60(SEDM)}


\software{MATLAB \citep{MATLAB:2020a},  
          SNID \citep{Blondin2007}, 
          DASH \citep{MuthukrishnaDASH},
          the GROWTH Marshal \citep{Kasliwal2019GM},
          Fritz (\url{https://github.com/fritz-marshal/fritz}).}

\appendix

\section{Loss function definitions}\label{sec:loss}
The mean square error (MSE) is the mean of the squared distances between the target variable ($y_i$) and predicted values ($y_i^p$):
\begin{equation}\label{eq:MSE}
MSE=\frac{1}{n}\sum_{i=1}^n(y_i-y_i^p)^2
\end{equation}
The mean bias error (MBE) is the mean of the distances between $y_i$ and $y_i^p$:
\begin{equation}\label{eq:MBE}
MBE=\frac{1}{n}\sum_{i=1}^n(y_i-y_i^p)
\end{equation}
The mean percentage error (MPE) is the mean of the distances between $y_i$ and $y_i^p$ divided by $y_i$:
\begin{equation}\label{eq:MPE}
MPE=\frac{1}{n}\sum_{i=1}^n\frac{(y_i-y_i^p)}{y_i}
\end{equation}
The mean absolute error (MAE) is the mean of the absolute distances between $y_i$ and $y_i^p$:
\begin{equation}\label{eq:MAE}
MAE=\frac{1}{n}\sum_{i=1}^n|y_i-y_i^p|
\end{equation}

\section{SNID and DASH ROC Curves}\label{sec:rocappendix}
To evaluate the performance of \texttt{SNID} (v5.0) we use the standard template bank included with the software\footnote{\url{https://people.lam.fr/blondin.stephane/software/snid/\#Download}}, the SNe Ia and non SN templates from the Berkeley SN Ia program (BSNIP; \citealt{bsnip}), the SN Ib and Ic templates from \citet{modjaz2014,modjaz2016,liu2016}, and  \citet{williamson}, and the SN IIP templates from \citet{gutirrez}. The ROC curve shown in Figure~\ref{fig:SNIascore} was computed by varying the minimal threshold for the \texttt{rlap} parameter between 0 and 25. For each automatic classification we take the best \texttt{SNID} match with the highest \texttt{rlap} value for that particular spectrum, as we have found that this performs better than considering multiple matches (e.g., counting how many SNe Ia matches are among the top 10 best matches). The redshift range was restricted to a reasonable range for the SNe expected to be found by the BTS ($z<0.2$). We place no restriction on the phase of the templates, and we have not attempted to restrict the template set used when running \texttt{SNID}. For future work it may be of interest to investigate if there is an optimal set of templates to use for SN Ia binary classification.

In order to evaluate \texttt{DASH} (v1.0 with Models\_v06) we used the default template bank included with the software\footnote{\url{https://github.com/daniel-muthukrishna/astrodash}}. To create the ROC curve we have investigated the effect of cuts on both the \texttt{softmax} scores and the \texttt{rlap} values that \texttt{DASH} produces. We investigate the ranges $\texttt{softmax}>0.2-1$ and $\texttt{rlap}>0-25$. To create the ROC curve shown in Figure~\ref{fig:SNIascore} we consider all possible cuts in the range $\texttt{softmax}>0.5-0.9$ and $\texttt{rlap}>0-25$, since for this range  we observe the most stable behavior in \texttt{DASH}. More extreme cuts on $\texttt{softmax}>0.92-0.99$ can offer some improvement on the BTS18 dataset, but only for one specific and narrow range of FPR (FPR~$=0.02$), which then gives TPR~$=0.62$. The rest of the ROC curve becomes noisy and with no real improvement over the one produced for $\texttt{softmax}>0.5-0.9$. As such, we consider this behavior to be unstable and do not recommend extreme cuts on the \texttt{softmax} score for SEDM data.



\bibliography{references}{}

\begin{thebibliography}{}
\expandafter\ifx\csname natexlab\endcsname\relax\def\natexlab#1{#1}\fi
\providecommand{\url}[1]{\href{#1}{#1}}
\providecommand{\dodoi}[1]{doi:~\href{http://doi.org/#1}{\nolinkurl{#1}}}
\providecommand{\doeprint}[1]{\href{http://ascl.net/#1}{\nolinkurl{http://ascl.net/#1}}}
\providecommand{\doarXiv}[1]{\href{https://arxiv.org/abs/#1}{\nolinkurl{https://arxiv.org/abs/#1}}}

\bibitem[{{Bellm} {et~al.}(2019{\natexlab{a}}){Bellm}, {Kulkarni}, {Graham},
  {Dekany}, {Smith}, {Riddle}, {Masci}, {Helou}, {Prince}, {Adams},
  {Barbarino}, {Barlow}, {Bauer}, {Beck}, {Belicki}, {Biswas}, {Blagorodnova},
  {Bodewits}, {Bolin}, {Brinnel}, {Brooke}, {Bue}, {Bulla}, {Burruss}, {Cenko},
  {Chang}, {Connolly}, {Coughlin}, {Cromer}, {Cunningham}, {De}, {Delacroix},
  {Desai}, {Duev}, {Eadie}, {Farnham}, {Feeney}, {Feindt}, {Flynn},
  {Franckowiak}, {Frederick}, {Fremling}, {Gal-Yam}, {Gezari}, {Giomi},
  {Goldstein}, {Golkhou}, {Goobar}, {Groom}, {Hacopians}, {Hale}, {Henning},
  {Ho}, {Hover}, {Howell}, {Hung}, {Huppenkothen}, {Imel}, {Ip}, {Ivezi{\'c}},
  {Jackson}, {Jones}, {Juric}, {Kasliwal}, {Kaspi}, {Kaye}, {Kelley},
  {Kowalski}, {Kramer}, {Kupfer}, {Landry}, {Laher}, {Lee}, {Lin}, {Lin},
  {Lunnan}, {Giomi}, {Mahabal}, {Mao}, {Miller}, {Monkewitz}, {Murphy},
  {Ngeow}, {Nordin}, {Nugent}, {Ofek}, {Patterson}, {Penprase}, {Porter},
  {Rauch}, {Rebbapragada}, {Reiley}, {Rigault}, {Rodriguez}, {van Roestel},
  {Rusholme}, {van Santen}, {Schulze}, {Shupe}, {Singer}, {Soumagnac}, {Stein},
  {Surace}, {Sollerman}, {Szkody}, {Taddia}, {Terek}, {Van Sistine}, {van
  Velzen}, {Vestrand}, {Walters}, {Ward}, {Ye}, {Yu}, {Yan}, \&
  {Zolkower}}]{Bellm2019}
{Bellm}, E.~C., {Kulkarni}, S.~R., {Graham}, M.~J., {et~al.}
  2019{\natexlab{a}}, \pasp, 131, 018002, \dodoi{10.1088/1538-3873/aaecbe}

\bibitem[{{Bellm} {et~al.}(2019{\natexlab{b}}){Bellm}, {Kulkarni}, {Barlow},
  {Feindt}, {Graham}, {Goobar}, {Kupfer}, {Ngeow}, {Nugent}, {Ofek}, {Prince},
  {Riddle}, {Walters}, \& {Ye}}]{Bellm2019b}
{Bellm}, E.~C., {Kulkarni}, S.~R., {Barlow}, T., {et~al.} 2019{\natexlab{b}},
  \pasp, 131, 068003, \dodoi{10.1088/1538-3873/ab0c2a}

\bibitem[{{Blagorodnova} {et~al.}(2018){Blagorodnova}, {Neill}, {Walters},
  {Kulkarni}, {Fremling}, {Ben-Ami}, {Dekany}, {Fucik}, {Konidaris}, {Nash},
  {Ngeow}, {Ofek}, {O' Sullivan}, {Quimby}, {Ritter}, \&
  {Vyhmeister}}]{Blagorodnova2018}
{Blagorodnova}, N., {Neill}, J.~D., {Walters}, R., {et~al.} 2018, \pasp, 130,
  035003, \dodoi{10.1088/1538-3873/aaa53f}

\bibitem[{{Blondin} \& {Tonry}(2007)}]{Blondin2007}
{Blondin}, S., \& {Tonry}, J.~L. 2007, in American Institute of Physics
  Conference Series, Vol. 924, The Multicolored Landscape of Compact Objects
  and Their Explosive Origins, ed. T.~{di Salvo}, G.~L. {Israel},
  L.~{Piersant}, L.~{Burderi}, G.~{Matt}, A.~{Tornambe}, \& M.~T. {Menna},
  312--321, \dodoi{10.1063/1.2774875}

\bibitem[{{Dahiwale} \& {Fremling}(2020)}]{ZTF20aatxryt}
{Dahiwale}, A., \& {Fremling}, C. 2020, Transient Name Server Classification
  Report, 2020-947, 1

\bibitem[{{Dekany} {et~al.}(2020){Dekany}, {Smith}, {Riddle}, {Feeney},
  {Porter}, {Hale}, {Zolkower}, {Belicki}, {Kaye}, {Henning}, {Walters},
  {Cromer}, {Delacroix}, {Rodriguez}, {Reiley}, {Mao}, {Hover}, {Murphy},
  {Burruss}, {Baker}, {Kowalski}, {Reif}, {Mueller}, {Bellm}, {Graham}, \&
  {Kulkarni}}]{Dekany20}
{Dekany}, R., {Smith}, R.~M., {Riddle}, R., {et~al.} 2020, \pasp, 132, 038001,
  \dodoi{10.1088/1538-3873/ab4ca2}

\bibitem[{Duev {et~al.}(2019)Duev, Mahabal, Masci, Graham, Rusholme, Walters,
  Karmarkar, Frederick, Kasliwal, Rebbapragada, {et~al.}}]{Duev2019rb}
Duev, D.~A., Mahabal, A., Masci, F.~J., {et~al.} 2019, Monthly Notices of the
  Royal Astronomical Society, 489, 3582

\bibitem[{{Fremling} \& {Sharma}(2018)}]{ZTF18aaxmhvk}
{Fremling}, C., \& {Sharma}, Y. 2018, Transient Name Server Classification
  Report, 2018-886, 1

\bibitem[{{Fremling} {et~al.}(2020){Fremling}, {Miller}, {Sharma}, {Dugas},
  {Perley}, {Taggart}, {Sollerman}, {Goobar}, {Graham}, {Neill}, {Nordin},
  {Rigault}, {Walters}, {Andreoni}, {Bagdasaryan}, {Belicki}, {Cannella},
  {Bellm}, {Cenko}, {De}, {Dekany}, {Frederick}, {Golkhou}, {Graham}, {Helou},
  {Ho}, {Kasliwal}, {Kupfer}, {Laher}, {Mahabal}, {Masci}, {Riddle},
  {Rusholme}, {Schulze}, {Shupe}, {Smith}, {Velzen}, {Yan}, {Yao}, {Zhuang}, \&
  {Kulkarni}}]{Fremling2020}
{Fremling}, C., {Miller}, A.~A., {Sharma}, Y., {et~al.} 2020, \apj, 895, 32,
  \dodoi{10.3847/1538-4357/ab8943}

\bibitem[{{Fremling} {et~al.}(2021){Fremling}, {Dahiwale}, {Mahabal}, {Miller},
  {Neill}, {Perley}, {Rigault}, {Sharma}, {Sollerman}, \&
  {Kulkarni}}]{astronote_SNIascore}
{Fremling}, C., {Dahiwale}, A., {Mahabal}, A., {et~al.} 2021, Transient Name
  Server Astronote, 2021-122

\bibitem[{{Graham} {et~al.}(2019){Graham}, {Kulkarni}, {Bellm}, {Adams},
  {Barbarino}, {Blagorodnova}, {Bodewits}, {Bolin}, {Brady}, {Cenko}, {Chang},
  {Coughlin}, {De}, {Eadie}, {Farnham}, {Feindt}, {Franckowiak}, {Fremling},
  {Gezari}, {Ghosh}, {Goldstein}, {Golkhou}, {Goobar}, {Ho}, {Huppenkothen},
  {Ivezi{\'c}}, {Jones}, {Juric}, {Kaplan}, {Kasliwal}, {Kelley}, {Kupfer},
  {Lee}, {Lin}, {Lunnan}, {Mahabal}, {Miller}, {Ngeow}, {Nugent}, {Ofek},
  {Prince}, {Rauch}, {van Roestel}, {Schulze}, {Singer}, {Sollerman}, {Taddia},
  {Yan}, {Ye}, {Yu}, {Barlow}, {Bauer}, {Beck}, {Belicki}, {Biswas}, {Brinnel},
  {Brooke}, {Bue}, {Bulla}, {Burruss}, {Connolly}, {Cromer}, {Cunningham},
  {Dekany}, {Delacroix}, {Desai}, {Duev}, {Feeney}, {Flynn}, {Frederick},
  {Gal-Yam}, {Giomi}, {Groom}, {Hacopians}, {Hale}, {Helou}, {Henning},
  {Hover}, {Hillenbrand}, {Howell}, {Hung}, {Imel}, {Ip}, {Jackson}, {Kaspi},
  {Kaye}, {Kowalski}, {Kramer}, {Kuhn}, {Landry}, {Laher}, {Mao}, {Masci},
  {Monkewitz}, {Murphy}, {Nordin}, {Patterson}, {Penprase}, {Porter},
  {Rebbapragada}, {Reiley}, {Riddle}, {Rigault}, {Rodriguez}, {Rusholme}, {van
  Santen}, {Shupe}, {Smith}, {Soumagnac}, {Stein}, {Surace}, {Szkody}, {Terek},
  {Van Sistine}, {van Velzen}, {Vestrand}, {Walters}, {Ward}, {Zhang}, \&
  {Zolkower}}]{Graham2019}
{Graham}, M.~J., {Kulkarni}, S.~R., {Bellm}, E.~C., {et~al.} 2019, \pasp, 131,
  078001, \dodoi{10.1088/1538-3873/ab006c}

\bibitem[{{Guti{\'e}rrez} {et~al.}(2017){Guti{\'e}rrez}, {Anderson}, {Hamuy},
  {Morrell}, {Gonz{\'a}lez-Gaitan}, {Stritzinger}, {Phillips}, {Galbany},
  {Folatelli}, {Dessart}, {Contreras}, {Della Valle}, {Freedman}, {Hsiao},
  {Krisciunas}, {Madore}, {Maza}, {Suntzeff}, {Prieto}, {Gonz{\'a}lez},
  {Cappellaro}, {Navarrete}, {Pizzella}, {Ruiz}, {Smith}, \&
  {Turatto}}]{gutirrez}
{Guti{\'e}rrez}, C.~P., {Anderson}, J.~P., {Hamuy}, M., {et~al.} 2017, \apj,
  850, 89, \dodoi{10.3847/1538-4357/aa8f52}

\bibitem[{Kasliwal {et~al.}(2019)Kasliwal, Cannella, Bagdasaryan, Hung, Feindt,
  Singer, Coughlin, Fremling, Walters, Duev, Itoh, \& Quimby}]{Kasliwal2019GM}
Kasliwal, M.~M., Cannella, C., Bagdasaryan, A., {et~al.} 2019, Publications of
  the Astronomical Society of the Pacific, 131, 038003,
  \dodoi{10.1088/1538-3873/aafbc2}

\bibitem[{{Kingma} \& {Ba}(2014)}]{ADAM}
{Kingma}, D.~P., \& {Ba}, J. 2014, arXiv e-prints, arXiv:1412.6980.
\newblock \doarXiv{1412.6980}

\bibitem[{{Kulkarni}(2020)}]{Kulkarni2020}
{Kulkarni}, S.~R. 2020, arXiv e-prints, arXiv:2004.03511.
\newblock \doarXiv{2004.03511}

\bibitem[{{Liu} {et~al.}(2016){Liu}, {Modjaz}, {Bianco}, \& {Graur}}]{liu2016}
{Liu}, Y.-Q., {Modjaz}, M., {Bianco}, F.~B., \& {Graur}, O. 2016, \apj, 827,
  90, \dodoi{10.3847/0004-637X/827/2/90}

\bibitem[{{Masci} {et~al.}(2019){Masci}, {Laher}, {Rusholme}, {Shupe}, {Groom},
  {Surace}, {Jackson}, {Monkewitz}, {Beck}, {Flynn}, {Terek}, {Landry},
  {Hacopians}, {Desai}, {Howell}, {Brooke}, {Imel}, {Wachter}, {Ye}, {Lin},
  {Cenko}, {Cunningham}, {Rebbapragada}, {Bue}, {Miller}, {Mahabal}, {Bellm},
  {Patterson}, {Juri{\'c}}, {Golkhou}, {Ofek}, {Walters}, {Graham}, {Kasliwal},
  {Dekany}, {Kupfer}, {Burdge}, {Cannella}, {Barlow}, {Van Sistine}, {Giomi},
  {Fremling}, {Blagorodnova}, {Levitan}, {Riddle}, {Smith}, {Helou}, {Prince},
  \& {Kulkarni}}]{Masci2019}
{Masci}, F.~J., {Laher}, R.~R., {Rusholme}, B., {et~al.} 2019, \pasp, 131,
  018003, \dodoi{10.1088/1538-3873/aae8ac}

\bibitem[{MATLAB(2020)}]{MATLAB:2020a}
MATLAB. 2020, 9.8.0.1359463 (R2020a) (Natick, Massachusetts: The MathWorks
  Inc.)

\bibitem[{{Modjaz} {et~al.}(2016){Modjaz}, {Liu}, {Bianco}, \&
  {Graur}}]{modjaz2016}
{Modjaz}, M., {Liu}, Y.~Q., {Bianco}, F.~B., \& {Graur}, O. 2016, \apj, 832,
  108, \dodoi{10.3847/0004-637X/832/2/108}

\bibitem[{{Modjaz} {et~al.}(2014){Modjaz}, {Blondin}, {Kirshner}, {Matheson},
  {Berlind}, {Bianco}, {Calkins}, {Challis}, {Garnavich}, {Hicken}, {Jha},
  {Liu}, \& {Marion}}]{modjaz2014}
{Modjaz}, M., {Blondin}, S., {Kirshner}, R.~P., {et~al.} 2014, \aj, 147, 99,
  \dodoi{10.1088/0004-6256/147/5/99}

\bibitem[{{Muthukrishna} {et~al.}(2019{\natexlab{a}}){Muthukrishna}, {Narayan},
  {Mandel}, {Biswas}, \& {Hlo{\v{z}}ek}}]{RAPID}
{Muthukrishna}, D., {Narayan}, G., {Mandel}, K.~S., {Biswas}, R., \&
  {Hlo{\v{z}}ek}, R. 2019{\natexlab{a}}, \pasp, 131, 118002,
  \dodoi{10.1088/1538-3873/ab1609}

\bibitem[{{Muthukrishna} {et~al.}(2019{\natexlab{b}}){Muthukrishna},
  {Parkinson}, \& {Tucker}}]{MuthukrishnaDASH}
{Muthukrishna}, D., {Parkinson}, D., \& {Tucker}, B.~E. 2019{\natexlab{b}},
  \apj, 885, 85, \dodoi{10.3847/1538-4357/ab48f4}

\bibitem[{{Nordin} {et~al.}(2019){Nordin}, {Brinnel}, {van Santen}, {Bulla},
  {Feindt}, {Franckowiak}, {Fremling}, {Gal-Yam}, {Giomi}, {Kowalski},
  {Mahabal}, {Miranda}, {Rauch}, {Reusch}, {Rigault}, {Schulze}, {Sollerman},
  {Stein}, {Yaron}, {van Velzen}, \& {Ward}}]{Nordin2019}
{Nordin}, J., {Brinnel}, V., {van Santen}, J., {et~al.} 2019, \aap, 631, A147,
  \dodoi{10.1051/0004-6361/201935634}

\bibitem[{{Perley} {et~al.}(2020){Perley}, {Fremling}, {Sollerman}, {Miller},
  {Dahiwale}, {Sharma}, {Bellm}, {Biswas}, {Brink}, {Bruch}, {De}, {Dekany},
  {Drake}, {Duev}, {Filippenko}, {Gal-Yam}, {Goobar}, {Graham}, {Graham}, {Ho},
  {Irani}, {Kasliwal}, {Kim}, {Kulkarni}, {Mahabal}, {Masci}, {Modak}, {Neill},
  {Nordin}, {Riddle}, {Soumagnac}, {Strotjohann}, {Schulze}, {Taggart},
  {Tzanidakis}, {Walters}, \& {Yan}}]{Perley2020rcf}
{Perley}, D.~A., {Fremling}, C., {Sollerman}, J., {et~al.} 2020, arXiv
  e-prints, arXiv:2009.01242.
\newblock \doarXiv{2009.01242}

\bibitem[{{Rigault} {et~al.}(2019){Rigault}, {Neill}, {Blagorodnova}, {Dugas},
  {Feeney}, {Walters}, {Brinnel}, {Copin}, {Fremling}, {Nordin}, \&
  {Sollerman}}]{Rigault2019}
{Rigault}, M., {Neill}, J.~D., {Blagorodnova}, N., {et~al.} 2019, \aap, 627,
  A115, \dodoi{10.1051/0004-6361/201935344}

\bibitem[{{Shappee} {et~al.}(2014){Shappee}, {Prieto}, {Stanek}, {Kochanek},
  {Holoien}, {Jencson}, {Basu}, {Beacom}, {Szczygiel}, {Pojmanski},
  {Brimacombe}, {Dubberley}, {Elphick}, {Foale}, {Hawkins}, {Mullins},
  {Rosing}, {Ross}, \& {Walker}}]{Shappee2014}
{Shappee}, B., {Prieto}, J., {Stanek}, K.~Z., {et~al.} 2014, in American
  Astronomical Society Meeting Abstracts, Vol. 223, American Astronomical
  Society Meeting Abstracts \#223, 236.03

\bibitem[{{Sherstinsky}(2020)}]{RNNreview}
{Sherstinsky}, A. 2020, Physica D Nonlinear Phenomena, 404, 132306,
  \dodoi{10.1016/j.physd.2019.132306}

\bibitem[{{Silverman} {et~al.}(2012){Silverman}, {Foley}, {Filippenko},
  {Ganeshalingam}, {Barth}, {Chornock}, {Griffith}, {Kong}, {Lee}, {Leonard},
  {Matheson}, {Miller}, {Steele}, {Barris}, {Bloom}, {Cobb}, {Coil},
  {Desroches}, {Gates}, {Ho}, {Jha}, {Kandrashoff}, {Li}, {Mandel}, {Modjaz},
  {Moore}, {Mostardi}, {Papenkova}, {Park}, {Perley}, {Poznanski}, {Reuter},
  {Scala}, {Serduke}, {Shields}, {Swift}, {Tonry}, {Van Dyk}, {Wang}, \&
  {Wong}}]{bsnip}
{Silverman}, J.~M., {Foley}, R.~J., {Filippenko}, A.~V., {et~al.} 2012, \mnras,
  425, 1789, \dodoi{10.1111/j.1365-2966.2012.21270.x}

\bibitem[{{SNIascore}(2021)}]{SN2021ijb_report}
{SNIascore}. 2021, Transient Name Server Classification Report, 9385

\bibitem[{{Tonry} {et~al.}(2018{\natexlab{a}}){Tonry}, {Stalder}, {Denneau},
  {Heinze}, {Weiland}, {Rest}, {Smith}, {Smartt}, {Young}, {Fulton}, {McBrien},
  {O'Neill}, \& {Clark}}]{SN2018cne}
{Tonry}, J., {Stalder}, B., {Denneau}, L., {et~al.} 2018{\natexlab{a}},
  Transient Name Server Discovery Report, 2018-818, 1

\bibitem[{{Tonry} {et~al.}(2020){Tonry}, {Denneau}, {Heinze}, {Weiland},
  {Flewelling}, {Stalder}, {Rest}, {Stubbs}, {Smith}, {Smartt}, {Young},
  {Srivastav}, {McBrien}, {O'Neill}, {Clark}, {Fulton}, {Gillanders}, {Dobson},
  {Chen}, {Wright}, \& {Anderson}}]{SN2020eyj}
{Tonry}, J., {Denneau}, L., {Heinze}, A., {et~al.} 2020, Transient Name Server
  Discovery Report, 2020-863, 1

\bibitem[{{Tonry} {et~al.}(2018{\natexlab{b}}){Tonry}, {Denneau}, {Heinze},
  {Stalder}, {Smith}, {Smartt}, {Stubbs}, {Weiland}, \& {Rest}}]{Tonry2018}
{Tonry}, J.~L., {Denneau}, L., {Heinze}, A.~N., {et~al.} 2018{\natexlab{b}},
  \pasp, 130, 064505, \dodoi{10.1088/1538-3873/aabadf}

\bibitem[{van~der Walt {et~al.}(2019)van~der Walt, Crellin-Quick, \&
  Bloom}]{Skyportal2019}
van~der Walt, S.~J., Crellin-Quick, A., \& Bloom, J.~S. 2019, Journal of Open
  Source Software, 4, \dodoi{10.21105/joss.01247}

\bibitem[{{Williamson} {et~al.}(2019){Williamson}, {Modjaz}, \&
  {Bianco}}]{williamson}
{Williamson}, M., {Modjaz}, M., \& {Bianco}, F.~B. 2019, \apjl, 880, L22,
  \dodoi{10.3847/2041-8213/ab2edb}

\end{thebibliography}
\bibliographystyle{aasjournal}

\end{document}